\documentclass[12pt,a4paper,final]{iopart}
\usepackage{iopams}  
\usepackage{graphicx}
\usepackage{dcolumn}  
\usepackage{bm}       
\usepackage{amssymb} 
\usepackage{braket}
\usepackage{mathcomp}

\usepackage{bbold}
\expandafter\let\csname equation*\endcsname\relax
\expandafter\let\csname endequation*\endcsname\relax 
\usepackage{amsmath}
\usepackage{physics}

\begin{document}

\title[Preparing an article for IOP journals in  \LaTeXe]{Eternal black holes and temporal quantum correlations}
\vspace{3pc}

\author{Ovidiu Racorean}
\address{General Direction of Information Technology}
\address{Banul Antonache str. 52-60, sc.C, ap.19, Bucharest, Romania}
\ead{ovidiu.racorean@mfinante.gov.ro}
\vspace{3pc}

\begin{abstract}
\vspace{1pc}

Recent works suggest that quantum theory may support a unification of the notions of space and time, thus treating the spatial and temporal quantum correlations equally. Specifically, the partial transposition of the maximally entangled state of two quantum systems at one time exactly matches the temporal quantum correlations of one quantum system that evolves unitary between two distinct moments of time. In this work we consider this equivalence of spatial and temporal quantum correlations in the context of AdS/CFT duality. We show that in the high temperature limit the thermofield double state is equivalent to the temporal quantum correlations of a quantum theory evolving unitary between two times. Thus, on the gravity side, we argue that the temporal correlations correspond to a black hole at one time connected behind the horizon by an Einstein-Rosen bridge to the same black hole at another time. Moreover, we construct the spacetime corresponding to this temporal wormholes and find that the spacetime is consistent with the interior solution of the BTZ black hole. We conclude suggesting that the BTZ interior solution could have an interpretation in the dS/CFT correspondence framework.

\end{abstract}

%Uncomment for PACS numbers title message
%\pacs{04.70.Dy, 03.67.Bg, 42.50.Ex, 95.30.Gv}
% Keywords required only for MST, PB, PMB, PM, JOA, JOB? 
\vspace{7pc}
\noindent{}      \hspace{7pc}          
\vspace{1pc}

\noindent{}       \hspace{10pc}             
\vspace{2pc}

\noindent{}  \hspace{15pc}  
% Uncomment for Submitted to journal title message
%\submitto{ Essay written for the Gravity Research Foundation 2017 Awards for Essays on Gravitation}
% Comment out if separate title page not required
\maketitle

\section{Introduction}

It was shown in recent works that quantum theory may support a unification of the notions of space and time, as such, treating spatial and time correlations equally.  Whether we talk about using a pseudo-density matrix formalism \cite{zhao}, \cite{zhang}, \cite{full} or a quantum generalization of Bayes’ theorem \cite{leifer}, all this attempts can be viewed as an extension of Jamiołkowski isomorphism \cite{mar}, \cite{marc}, \cite{marco} that maps the spatial quantum correlation of two distinct quantum systems at one time to temporal quantum correlations that involve a single quantum system at two different times. Accordingly, the spatial correlations were found to exactly correspond to temporal correlations via partial transposition. To be more precise, the partial transposition of the maximally entangled state of two quantum systems exactly matches the correlations of one quantum system that evolves unitary between two distinct moments of time. 

We should emphasize here that this equivalence implies a slight modification of traditional quantum theory. As was explicitly argued in \cite{leifer}, the quantum system measured at two times is defined on two distinct Hilbert spaces, i.e. two different quantum systems, such that the temporal correlations are defined on a tensor product of two Hilbert spaces.  As a result, we can say that the Hilbert space of spatial correlations is equivalent with Hilbert space of temporal correlations.

In this work we extend this line of thinking to the AdS/CFT correspondence. Thus, we argue that in the high-temperature limit the thermofield double state \cite{mal}, \cite{suss}, \cite{raam}, i.e. the space correlations of two spatially separated CFT’s at one time, is equivalent to temporal correlations of one CFT evolving unitary between two times or, equivalently, of two different CFTs temporally separated. Consequently, on the gravity side of the AdS/CFT duality the temporal correlations should correspond to one black hole at two different times or two distinct temporally separated black holes. Furthermore, we assume that the duality holds for the temporal correlations as well such that the two temporally separated black holes are connected behind the horizon by an Einstein-Rosen bridge. We may see the spatial wormhole as an Einstein-Rosen bridge connecting two spatially separated black holes on the same space-like hypersurface while the temporal wormhole as an Einstein-Rosen bridge connecting two temporally separated black holes on two different space-like hypersurfaces. To this end we can consider the two types of wormholes (spatial and temporal) as being equivalent when the roles of space and time coordinates are interchanged.

At this point we should ask what the corresponding spacetime of these unusual temporal wormholes should be. The construction of the spacetime of temporal wormholes is related to the remark that space and time coordinates interchange their roles in the metric of the corresponding BTZ black holes. Since such interchange of the spatial and temporal roles occurs behind the event horizon we assume that the corresponding spacetime of temporal wormholes should have its origin on the interior BTZ solution \cite{breh}, \cite{doran}, \cite{barr}, \cite{naka}. Accordingly, we further derive the line element of the BTZ black holes for the special case of switching the roles of space and time coordinates. The metric we derive resembles a static de Sitter space. 

In a series of papers \cite{das}, \cite{nar}, \cite{nara}, \cite{naray} a similar form of the metric is discussed in relation to dS/CFT duality. Thus, our result could lead to the conclusion that spacetime corresponding to the BTZ interior solution could have an interpretation in the dS/CFT correspondence framework. In this context the gravity dual of the quantum temporal correlations is a static de Sitter space. The temporal wormhole connecting two temporally separated black holes is a de Sitter spacetime. 

Although, numerous aspects remain unclear, such as the traversability of these peculiar temporal wormholes, the present work may prove to be important at the conceptual level.

\section{Equivalence of space correlations and time correlations}

In quantum theory space and time are not treated on an equal footing. In this sense, the maximally entangled state of two quantum systems at one time is treated differently with respect to one unitary evolving system at two times. The maximally entangled state is described by a tensor product of two Hilbert spaces while the single state at two times is described by a single Hilbert state and a dynamical map between the input and output states. 

However, in recent works \cite{zhao}, \cite{zhang}, \cite{leifer}, \cite{full}, \cite{mar}, \cite{marc}, \cite{marco} we observe a tendency to bring these two separate quantum descriptions under the same conceptual umbrella. These attempts consider a slightly modified Hilbert space formalism in the sense that pairs of systems spacelike separated and single systems at two different times (temporally separated) are both described on a tensor product of two Hilbert spaces. In addition, although spatial and temporal correlations are described by operators on the tensor product, they differ from one another by a partial transpose. One other way to express this equivalence is to consider that a maximally entangled state equals a maximally mixed state that evolves unitary between two times.

Let us now explore this concept in greater depth. To do this we begin considering the case of a pure state $\ket{\Psi}$  with $\rho=\ket{\Psi}\bra{\Psi}=\frac{1}{d}\sum_{n,j}\ket{n}\ket{n}\bra{j}\bra{j}$  of two quantum systems, $L$ and $R$, defined on the Hilbert space $\mathcal{H}_{LR}=\mathcal{H}_R\otimes \mathcal{H}_L$ such that:

\begin{equation}
\ket{\Psi}=\frac{1}{d}\sum_n\ket{n}\ket{n} .
\end{equation}

The first step in finding the analogue of the maximally entangled state in Eq.(1) in the temporal context we should write down the partial transpose $\rho^{T_L}$ that represents the quantum state of temporal correlations. Thus, using the pure state $\rho$ we can express the partial transpose as:

\begin{equation}
\rho^{T_L}=\ket{\Psi}\bra{\Psi}^{T_L}=\frac{1}{d}\sum_{n,j}\ket{n}\ket{j}\bra{j}\bra{n} .
\end{equation}

It has been argued \cite{zhao} , \cite{zhang}, \cite{full}, \cite{leifer}, \cite{mar}, \cite{marc}, \cite{marco} that precisely this state equals the temporal correlated state, denoted $\rho^t$. As a result, we can emphasize that:

\begin{equation}
\rho^t=\rho^{T_L} .
\end{equation}

Since the density matrix $\rho^{T_L}$ is defined on the Hilbert space $\mathcal{H}_R\otimes \mathcal{H}_L$, the formalism of quantum theory is slightly modified in the sense that the state of temporal correlations should also be defined on a similar tensor product of two Hilbert spaces. Thus, the single quantum system evolving unitary is mapped at time $t_1$ to a Hilbert space $\mathcal{H}_i$ while at time $t_2$ it is mapped to another Hilbert space $\mathcal{H}_f$ such that the temporal quantum correlated state $\rho^t$ is defined on the tensor product $\mathcal{H}_{if}=\mathcal{H}_i\otimes \mathcal{H}_f$ .  In this context, an interesting result that Eq.(3) suggests is that we have the following equivalence:

\begin{equation}
\mathcal{H}_i\otimes \mathcal{H}_f=\mathcal{H}_R\otimes \mathcal{H}_L
\end{equation}

This equivalence suggests that the same Hilbert space can be regarded as the Hilbert space of two entangled subsystems or as the Hilbert space of a single quantum system evolving unitary between two times.

While we set up the equivalence between spatial and temporal correlations under a partial transpose we should stresses the meaning of the state $\rho^t$.  Consequently, the state $\rho^t$ should be seen as temporal correlations of a single quantum system $R$(or $L$) that evolves unitary between two measured instances of time, $t_1$ and $t_2$, with $t_1<t_2$. As a result, the measurement of the initial state by the operator $\mathcal{O}_i$ equals an operator acting on the right side, $\mathcal{O}_R$, while the operator acting on the final state $\mathcal{O}_f$ is equivalent to the transpose of the operator acting on the left side of the maximally entangled state: 

\begin{equation}
\mathcal{O}_i=\mathcal{O}_R, \mathcal{O}_f=\mathcal{{O}_L}^T .
\end{equation}

In other words, the temporal equivalent of the maximally entangled state $\rho$ (taken as a partial transpose) is the maximally mixed state $\rho_R$(or $\rho_L$) measured at two different times under the identity (unitary) evolution. To find the state of the system $R$ we take the partial trace of $\rho^{T_L}$, such that we have:

\begin{equation}
\rho_R=\frac{1}{d}\sum_n\ket{n}\bra{n} ,
\end{equation}

which is the maximally mixed state,$\rho_R=\frac{1}{d}\mathbb{1}$ . We should consider this state as the initial state of the temporal evolution while the final state would be the transpose of $\rho_L$: 

\begin{equation}
\rho_i=\rho_R,     \rho_f=\rho_L^T .
\end{equation}

We can consider that temporal correlation relates the two maximally mixed states $\rho_R$ and $\rho_L^T$ which are temporally separated by an interval, $[t_1, t_2]$.   

As shown in \cite{san}, the only physical interpretation of transpose is time reversal. In this respect, the partial transposition can be regarded as a partial time reversal in one of the two maximally entangled quantum systems. We would like to perform a reversal of time on the subsystem L to evaluate the quantum state that results. Starting from the initial quantum state $\rho$  such a partial time reversal in the system $L$ would lead us to the state $\rho^*$, that is:

\begin{equation}
\rho^*=\frac{1}{d}\sum_{n,j}\ket{n}\ket{n^*}\bra{j}\bra{j^*} ,
\end{equation}

which can be further arranged , as:

\begin{equation}
\rho^*=\frac{1}{d}\sum_{n,j}\ket{n}\ket{j}\bra{j}\bra{n} .
\end{equation}

Comparing with the state in Eq.(2), it can be easily noted that, $\rho^*=\rho^{T_L}=\rho^t$. The consequences of this identification are important in what follows and can be synthesised in the statement that the partial transpose density matrix, $\rho^{T_L}$ corresponds to the maximally entangled state:

\begin{equation}
\ket{EPR}=\frac{1}{d}\sum_{n,j}\ket{n}\ket{n^*} .
\end{equation}

That is to say, we can consider the vector state $\ket{EPR}$ as being equivalent to the vector state of temporal correlations. We can verify one more time this statement since we can infer from the  $\ket{EPR}$ vector state that an operator $\mathcal{O}_L$ acting on the left side is equivalent to its transpose acting on the right side:

\begin{equation}
\mathcal{O}_L\ket{EPR}=\mathcal{{O}_R}^T\ket{EPR} ,
\end{equation}

which is the same result as in the case of temporal correlation in Eq.(5).

The importance of the EPR state will become clear as we will try to extend the equivalence of spatial and temporal correlations to thermal states. 

\section{The thermofield double state and spatial wormholes}

We would like to extend the equivalence of spatial and temporal correlations in a key point of intersection between quantum theory and General Relativity, which is the AdS/CFT correspondence.  The motivation that we have in mind when considering this extension is twofold. First, we were motivated by the close link between space and time provided by the AdS/CFT duality.

The other important reason we chose to probe the equivalence of space and time correlations in the AdS/CFT realm is the form of the thermofield double state. The meaning of this statement is that the EPR state is closely related to the thermofield double state,

\begin{equation}
\ket{TFD}=\frac{1}{\sqrt{Z}}\sum_ne^\frac{-\beta E_n} {2}\ket{n}\ket{n^*} ,
\end{equation}

with the known notations, $Z$ as the partition function and $\beta$ as the inverse temperature. Here we consider two non-interacting copies of conformal field theory as two quantum systems, $L$ and $R$, such that we can decompose the Hilbert space $\mathcal{H}_{LR}$ of the composite system as $\mathcal{H}_{LR}=\mathcal{H}_R\otimes \mathcal{H}_L$.

Notably, the quantum system $L$ suffers a reversal of time, such that both CFTs evolve in the same direction of time. The motivation for this choice of time evolution in the quantum system $L$, should be identified on the gravity side of the AdS/CFT duality.  It has been argued \cite{mal} , \cite{suss}, \cite{raam} that precisely the TFD state is dual to two black holes whose interiors are connected by an Einstein-Rosen bridge. Accordingly, an observer in either asymptotic region sees the BTZ black hole spacetime, which is understood \cite{witt} to correspond to the thermal state of the conformal field theory. Thus, the motivation of choosing to take a reversal of time in the left system such that both CFTs evolve in the same direction of time  is translated on the gravity side as both black holes dual to CFT’s evolve in the same direction of time, as is sketched in Figure 1.

\begin{figure}
\includegraphics[width=7.6cm]{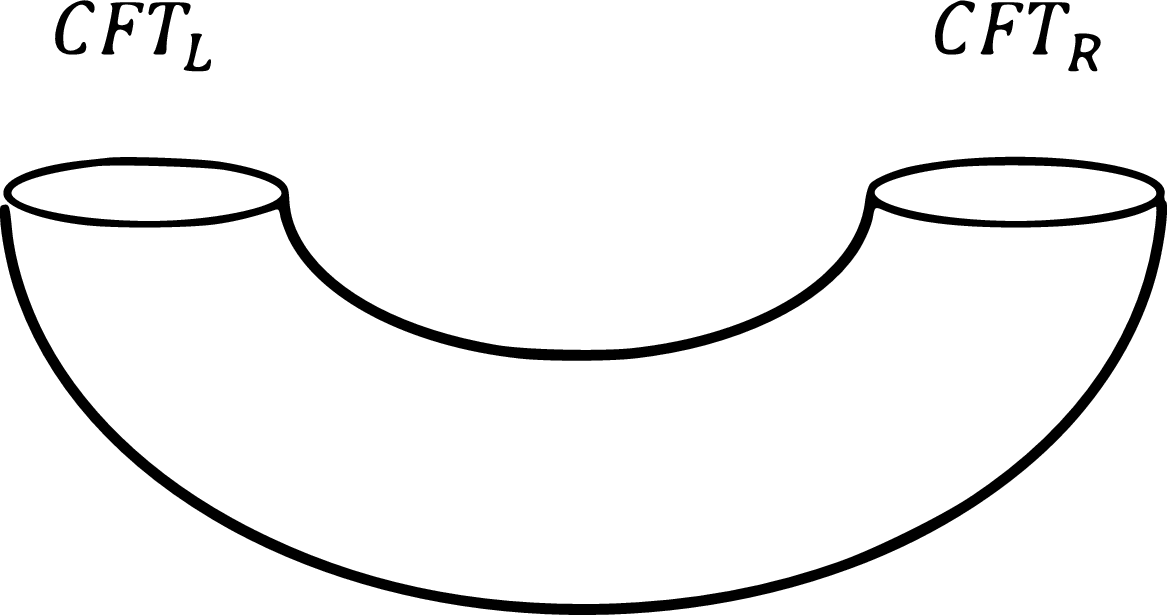}
\caption{\label{fig:fig1} The spatial Einstein-Rosen bridge connects two distant black holes on the same spacelike hypersurface.}
\end{figure}

We can consider this construction as a spatial wormhole that connects two spatially distant black holes across the same space hypersurface at a constant time. 

 Let us return to the TFD state, which as we have seen is precisely of the form of the EPR state with one system experiencing a time reversal. A comparison of Eq.(10) and Eq.(12)  clearly reveals that at finite temperatures we have:

\begin{equation}
\ket{TFD}=\sqrt{d\rho_R}\ket{EPR}=\sqrt{d{\rho_L}^T}\ket{EPR} ,
\end{equation}

 such that in the high-temperature limit ($\beta\longrightarrow0$) the two states are equal. 

We would like to define the temporal correlations within the limit of AdS/CFT duality. That is, we have to search for the temporal correlated state and for that we consider the density matrix of the TFD state vector:

\begin{equation}
\rho_{TFD}=\frac{1}{Z}\sum_{n,j}e^\frac{-\beta E_n}{2}\ket{n}\ket{n^*}\bra{j}\bra{j^*} .
\end{equation}

At high-temperatures, this state equals the partially transposed state $\rho^{T_L}$. As we recall, precisely this state is equivalent to the temporal correlated state $\rho^t$, such that we have, $\rho_{TFD}=\rho^t$. 

In our scenario, the TFD state in Eq.(12) is one part of the equivalence and it defines the spatial correlations between two spatially distant CFT systems, $R$ and $L$. Thus, on the other side of the equivalence, correlations of the thermal state of one CFT system that evolves unitary between measurements at two different instances of time should reside. To find the thermal state of the right CFT, we trace over the degrees of freedom of the left CFT to find the reduced density matrix:

\begin{equation}
\rho_R= Tr_L \ket{TFD}\bra{TFD} = \frac{1}{Z}\sum_ne^{-\beta E_n}\ket{n}\bra{n} ,
\end{equation}

which is exactly the thermal state of the right CFT,  

\begin{equation}
\rho_R=  \frac{1}{Z}\sum_ne^{-\beta E_n}\mathbb{1} .
\end{equation}

The reduced density matrix $\rho_R$ describes a maximally mixed state as required. That is, the thermofield double state, i.e., the maximally entangled state between two copies of quantum theory at one time, is equivalent in the high-temperature limit,  to the thermal state (maximally mixed state) of one CFT evolving unitary between two times.  Now, we can take $\rho_R$ as the initial state at time $t_1$ described by Hilbert space $\mathcal{H}_i$ and $\rho_L$ as the final state at time  $t_2$ described by the Hilbert space $\mathcal{H}_f$ as two temporally separated CFTs on the tensor product $\mathcal{H}_{if}=\mathcal{H}_i\otimes \mathcal{H}_f$. In this case, the temporal quantum correlated state relates the two temporally distant thermal states. Since we have seen that we have the equivalence $\mathcal{H}_i\otimes \mathcal{H}_f=\mathcal{H}_R\otimes \mathcal{H}_L$,  we can say that the tensor product can also be $\mathcal{H}_{LR}=\mathcal{H}_R\otimes \mathcal{H}_L$. In this case we can say that the same Hilbert space can be regarded as the Hilbert space of two entangled CFTs or as the Hilbert space of a single CFT evolving unitary between two times. We add here that a similar result can be found in \cite{bala}, in relation to de Sitter holography. We return to the idea of de Sitter involvement in our description of the temporal Einstein-Rosen bridge later.

On the gravity side, the reduced density matrix $\rho_R$ describes a maximally mixed state which in the high temperature limit corresponds to the BTZ black hole \cite{witt}. The gravity dual should be a BTZ black hole that evolves unitary between two times. We consider $\rho_R$ and $\rho_L$ which describe the same black hole at two distinct times as two different temporally separated black holes. 

If we assume that the AdS/CFT duality holds for the temporal correlations since they are equivalent to the TFD state , we should conclude that the two temporally separated black holes should be connected behind the horizon. We depict this result in Figure 2.

\begin{figure}
\includegraphics[width=7.6cm]{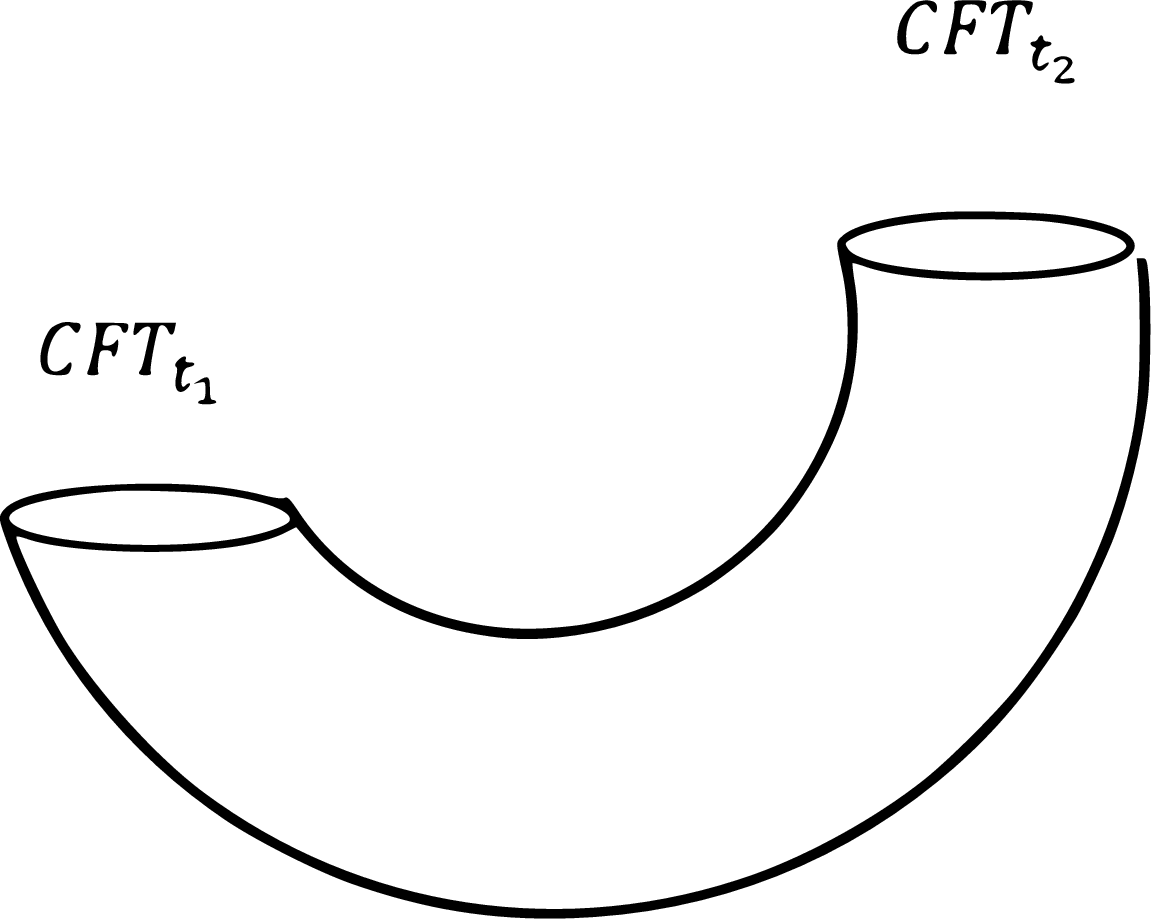}
\caption{\label{fig:fig2} . The temporal Einstein-Rosen bridge connecting a black hole on two different spacelike hypersurfaces.}
\end{figure}

Thus, the black hole at one time is connected behind the horizon to the same black hole at another time by an Einstein-Rosen bridge. We can imagine the temporal wormholes as connecting two black holes across different spacelike hypersurfaces at constant space coordinates. 

\section{Temporal wormholes}

We have seen that we can distinguish spatial wormholes from temporal wormholes. The spatial wormholes describe an Einstein-Rosen bridge connecting two spatially separated black holes while the temporal wormholes describe the connection under the horizon of two temporally separated black holes. 
On the quantum theory side of the AdS/CFT correspondence we have seen that TFD state is equivalent to temporal correlations.  The question to ask at this point is whether there is also equivalence between the two wormholes. If the answer is positive then we should be able to consider the temporal wormhole as eternal black hole in the Penrose diagram. In other words, can the temporal Einstein-Rosen bridge correspond to the Penrose diagram of the BTZ black hole solution?  

One way to view the temporal correlations as a solution of the BTZ black hole is to observe that the space and time roles are interchanged in our construction of the temporal wormholes. The two spatially separated black holes are connected by an Einstein-Rosen bridge on the same spacetime sheet in the case of spatial wormholes. In the case of the temporal wormholes the exact opposite is in place. The two temporally separated black holes are connected on two different spacetime hypersurfaces. As such, time and space coordinates suffer an inversion of roles.

This situation is similar to considering the inversion of space and time roles behind the horizon, on the interior of the black hole. Although the interior solution of black holes has a long history \cite{breh}, it has only recently been found to have some support \cite{doran}, \cite{barr}, \cite{naka}. The change that occurs in the nature of spacetime for the interior solution is exactly the interchange of the space and time coordinate characteristics.

Let us now elaborate further on this scenario and derive the line element of 2+1 dimensional BTZ black hole spacetime, assuming the interchanging of space and time roles. Thus, we start by considering a general radially symmetric line element:

\begin{equation}
ds^2=-B^2(t)dt^2+A^2(t)dr^2+F^2(t)d\phi^2.
\end{equation}

Taking into account the spherical symmetry and assuming that t is now the radial coordinate, we rewrite the metric as simply:

\begin{equation}
ds^2=-B^2(t)dt^2+A^2(t)dr^2+t^2d\phi^2
\end{equation}

and solve for the functions $B(t)$ and $A(t)$, respectively.
Indeed, after we determine the curvature two-form and solve the equation for the empty space with the cosmological constant $\Lambda$ (see the APPENDIX ) we recover the functions $A$ and $B$ as:

\begin{equation}
B^2(t) =\frac{1}{M-\frac{t^2}{\ell^2}} .
\end{equation}

and

\begin{equation}
A^{2}(t)=M-\frac{t^2}{\ell^2}.
\end{equation}

With these results at hand, we are now able to write the metric of the 2+1 black hole as:

\begin{equation}
ds^2=-\frac{1}{\left(M-\frac{t^2}{\ell^2}\right)}dt^2+\left(M-\frac{t^2}{\ell^2}\right)dr^2+t^2d\phi^2
\end{equation}

which is only valid for values  $\frac{t^2}{\ell^2}<M$.
Let us consider set for simplicity $M=\ell=1$ such that we remain with the line element:

\begin{equation}
ds^2=-\frac{1}{ 1-t^2 }dt^2 + \left(1-t^2 \right)dr^2+t^2 d\phi^2 
 \end{equation}

As such, the line element describes the region that refers to the exterior of the black hole spacetime. However, the metric in Eq.(22) describes this region as being a gravitationally trapped region in stark contrast with the exterior BTZ solution. Notably, when the roles of space and time are interchanged, the exterior region is mapped to a trapped region. 
We emphasize here that a similar form of the metric is discussed in relation to dS/CFT duality in \cite{das}, \cite{nar}, \cite{nara}, \cite{naray}. In what follows, we keep the line of thought in these papers. Accordingly, one important observation is that the spacetime in Eq.(22) has a Penrose diagram, as shown in Figure 3, which resembles that of the BTZ black hole rotated by $\frac{\pi}{2}$ . 

\begin{figure}
\includegraphics[width=8.6cm]{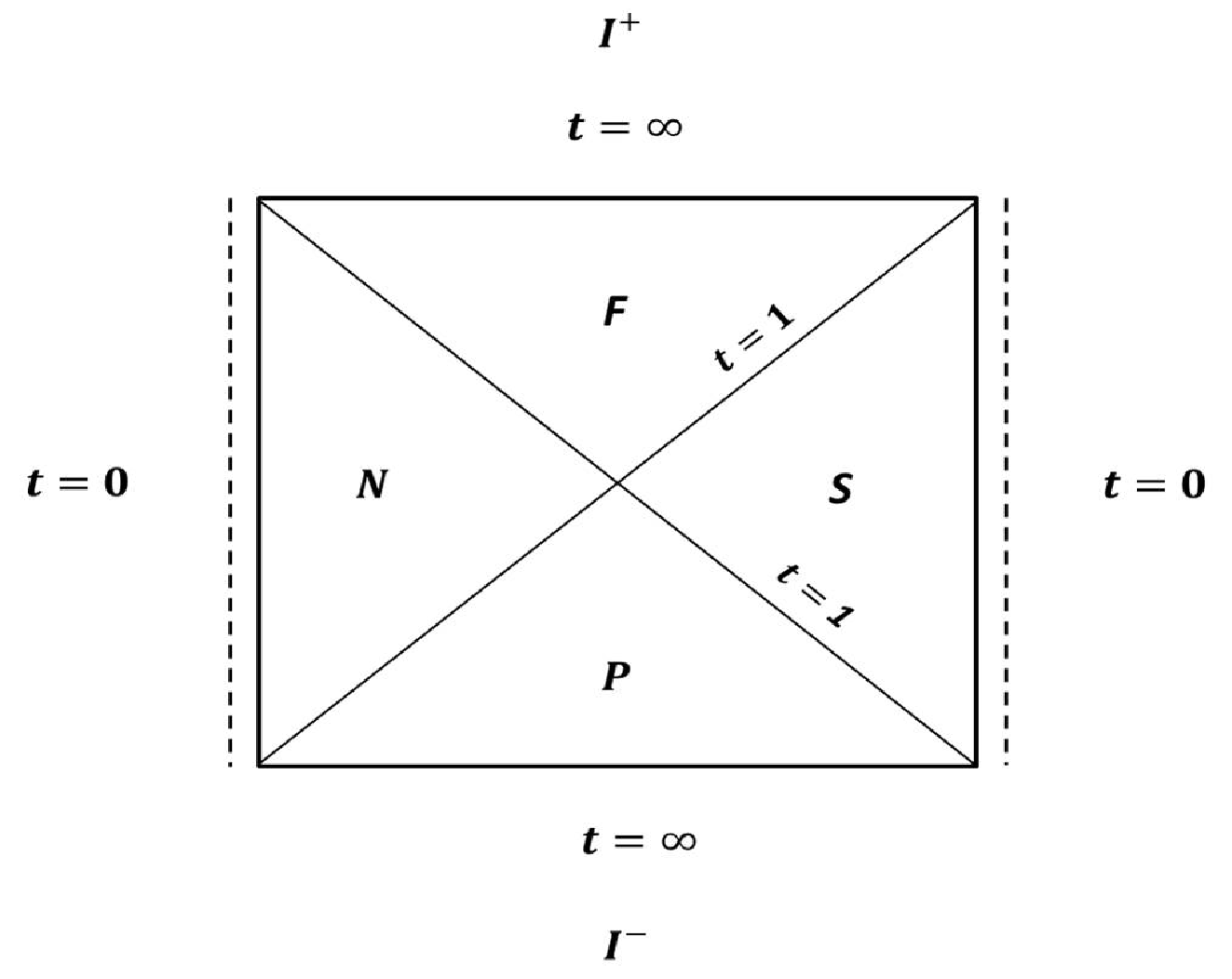}
\caption{\label{fig:fig3}. Penrose diagram of the interior solution of the BTZ black hole. }
\end{figure}

The metric in Eq.(22) is singular at the cosmological horizon $t=1$. There are two regions with $0\le t\le1$    which correspond to the causal diamonds of observers at the north (N) and south poles (S). In other words, an observer at  $t=0$ is surrounded by a cosmological horizon at $t=1$. Moreover, as shown in Figure 3, we also have two regions with $1\le t<\infty$ containing the future-null infinity $I^+$ and past-null infinity $I^-$, i.e., the future universe (F) and the past universe (P), respectively. It is easy to observe that this construction is similar to de Sitter spacetime.
The temporal quantum entanglement that occurs when a CFT is measured at two different times is dual to a spacetime that resembles the dS spacetime. Thus, this spacetime could have an interpretation in the dS/CFT correspondence framework.

\section{Conclusions}

Recent attempts to unify the quantum mechanics notions of space and time have led to the conclusion that spatial and temporal correlations should be treated equally. As such, the EPR state of two quantum systems at one time exactly equals the temporal correlations of one quantum system that evolves unitary between two distinct moments of time.

In this paper we took advantage of this quantum equivalence and we have extended this equivalence to the AdS/CFT duality framework. Accordingly, we have argued that in the high-temperature limit the TFD state is equivalent to temporal correlations of one CFT measured at two times that evolves unitary. 

We have shown that this equivalence has some important implications for the gravity side of the AdS/CFT duality. In this sense, we considered that the temporal correlations of one CFT at two times are dual to two black holes (temporally separated) connected behind the horizon by an Einstein-Rosen bridge. 

Furthermore, we have argued that these atypical temporal wormholes are related to the interior solution of the BTZ black hole. We derived the interior metric of the BTZ black holes and briefly discussed some intriguing implications of the identification between temporal wormholes and the interior solution.

\vspace{7pc}

\renewcommand{\theequation}{A-\arabic{equation}}
  % redefine the command that creates the equation no.
  \setcounter{equation}{0}  % reset counter 
  \section*{APPENDIX : Derivation of the interior BTZ metric }

To derive the interior BTZ metric we start by considering a general radially symmetric line element:

\begin{equation}
ds^2=-B^2(t)dt^2+A^2(t)dr^2+F^2(t)d\phi^2.
\end{equation}

Then, taking in account the spherical symmetry, and since we assumed that t is the radial coordinate we rewrite the metric as simply:

\begin{equation}
ds^2=-B^2(t)dt^2+A^2(t)dr^2+t^2d\phi^2
\end{equation}

and we attempt to derive the functions  $B(t)$ and $A(t)$. 
The metric should describe spacetime as seen by an observer situated behind the event horizon such that it would satisfy Einstein’s equations. Thus, we can write the vielbeins: 

\begin{equation}
e^0=B(t)dt,\     \  e^1=A(t)dr,\      \   e^2=td\phi.
\end{equation}

The components of the spin connection can be found via Cartan’s first structure equation $de^{\mu}=-\omega^{\mu}_{\nu}\wedge e^{\mu}$  as:

\begin{equation}
de^0=B^{'}(t)dt\wedge dt=0,
\end{equation}

since $dt\wedge dt=0$.

Furthermore, we identify

\begin{equation}
de^1=A^{'}(t)dt\wedge dr,
\end{equation}

and

\begin{equation}
de^2= dt\wedge d\phi.
\end{equation}

Now, we may infer that the only term with $dt \wedge dr$ is $\omega^{2}_{r} dr \wedge e^0$, such that we have: 

\begin{equation}
A^{'}(t)dt\wedge dr=-\omega^{2}_{r} dr \wedge B(t)dt,
\end{equation}

Thus, we can write 

\begin{equation}
\omega^{2}_{r} = -\frac{ A^{'}(t)}{ B(t)},
\end{equation}

Similarly, observing that the only term with $dt \wedge d\phi$ is $\omega^{1}_{\phi} d\phi \wedge e^0$, we conclude that 

\begin{equation}
dt\wedge d\phi=\omega^{1}_{\phi} d\phi \wedge B(t)dt,
\end{equation}

and further 

\begin{equation}
\omega^{1}_{\phi} = -\frac{1}{B(t)}.
\end{equation}

All the other terms vanished.

Thus, to summarize, we have

\begin{equation}
\omega^0=0,\     \  \omega^1=-\frac{1}{ B(t)},\      \   \omega^2=-\frac{ A^{'}(t)}{ B(t)}.
\end{equation}

Now we will find the curvature two-form  $R^{\mu}=d\omega^{mu}+\frac{1}{2}\omega^{\mu}_{\nu} \wedge \omega^{nu}$  and solve the equation for the empty space  $R^{\mu}=\frac{\Lambda}{2} e^{\mu}_{\nu} \wedge e^{\nu}$  with the cosmological constant $\Lambda$. 

To find $A(t)$ and $B(t)$ we need only two of the three equations that result:

\begin{equation}
R^0 = -\frac{A^{'}(t)}{ B^2(t)} dr \wedge d\phi=\Lambda A(t) t dr \wedge d\phi.
\end{equation}

and
\begin{equation}
R^1 = \frac{B^{'}(t)}{ B^2(t)} dt \wedge d\phi=-\Lambda B(t) t dt \wedge d\phi.
\end{equation}

since the equation for $R^1$ is only in terms of $B(t)$. 
We can form the differential equations

\begin{equation}
\begin{split}
\frac{A^{'}(t)}{B^2(t)}=-\Lambda A(t)t \\
\frac{B^{'}(t)}{B^2(t)}=-\Lambda B(t)t 
\end{split}
\end{equation}

Let us integrate the last differential equation with the separation of variables: 

\begin{equation}
\frac{B^{'}(t)}{ B^3(t)} =-\Lambda t.
\end{equation}

which yields:

\begin{equation}
-\frac{1}{ B^2(t)} =-\left(\Lambda t^2+M \right).
\end{equation}

where we set $M$ as the constant of integration. Thus we have:

\begin{equation}
B^2(t) =\frac{1}{\Lambda t^2+M} .
\end{equation}
 
If we take the AdS spacetime cosmological constant $\Lambda =-\frac{1}{\ell^2}$ we remain with:

\begin{equation}
B^2(t) =\frac{1}{M-\frac{t^2}{\ell^2}} .
\end{equation}

Now replacing $B^2 (t)$ in eq.(A-14) we find

\begin{equation}
\frac{A^{'}(t)}{A(t)}=\frac{-\Lambda t}{\Lambda t^2+M } .
\end{equation}

which by integration results in: 

\begin{equation}
A^{2}(t)=\Lambda t^2+M,
\end{equation}

or substituting the cosmological constant $\Lambda$ yields:

\begin{equation}
A^{2}(t)=M-\frac{t^2}{\ell^2}.
\end{equation}

With the help of eq.(A-18) and eq.(A-21) we can write the full metric of the 2+1 black hole

\begin{equation}
ds^2=-\frac{1}{\left(M-\frac{t^2}{\ell^2}\right)}dt^2+\left(M-\frac{t^2}{\ell^2}\right)dr^2+t^2d\phi^2
\end{equation}

which is only valid for values  $\frac{t^2}{\ell^2}<M$. As such, the line element describes the region that refers to the exterior of the black hole spacetime. However, the metric in Eq.(A-22)  describes this region as being a gravitationally trapped region in stark contrast with the exterior BTZ solution. Notably, when the roles of space and time are interchanged, the exterior region is mapped to a trapped region.

\end{document}